\documentclass[8.5pt,twoside,twocolumn]{article}
\usepackage[super,sort&compress,comma]{natbib} 
\usepackage{url}
\usepackage{mhchem}
\usepackage{times,mathptmx}
\usepackage{bm}
\usepackage{sectsty}
\usepackage{balance}
\usepackage{graphicx} 
\usepackage{lastpage}
\usepackage[format=plain,justification=raggedright,singlelinecheck=false,font=small,labelfont=bf,labelsep=space]{cap
tion} 
\usepackage{fancyhdr}
\usepackage{multirow}
\usepackage{float}

\begin{document}
\thispagestyle{plain}
\fancypagestyle{plain}{
\renewcommand{\headrulewidth}{1pt}}
\renewcommand{\thefootnote}{\fnsymbol{footnote}}
\renewcommand\footnoterule{\vspace*{1pt}%
\hrule width 3.4in height 0.4pt \vspace*{5pt}} 
\setcounter{secnumdepth}{5}
\makeatletter 
\def\subsubsection{\@startsection{subsubsection}{3}{10pt}{-1.25ex plus -1ex minus -.1ex}{0ex plus 0ex}{\normalsize\bf}} 
\def\paragraph{\@startsection{paragraph}{4}{10pt}{-1.25ex plus -1ex minus -.1ex}{0ex plus 0ex}{\normalsize\textit}} 
\renewcommand\@biblabel[1]{#1}            
\renewcommand\@makefntext[1]%
{\noindent\makebox[0pt][r]{\@thefnmark\,}#1}
\makeatother 
\renewcommand{\figurename}{\small{Fig.}~}
\sectionfont{\large}
\subsectionfont{\normalsize} 
\def\rr{{{\bf r}}}
\def\EE{{{\bf E}}}
\def\RR{{{\bf R}}}
\fancyfoot{}
\fancyhead{}
\renewcommand{\headrulewidth}{1pt} 
\renewcommand{\footrulewidth}{1pt}
\setlength{\arrayrulewidth}{1pt}
\setlength{\columnsep}{6.5mm}
\setlength\bibsep{1pt}

\twocolumn[
  \begin{@twocolumnfalse}
\noindent\LARGE{\textbf{Theoretical description of the efficiency enhancement in DSSC sensitized by newly synthesized heteroleptic Ru complexes }}
\vspace{0.6cm}

\noindent\large{\textbf{Yavar T. Azar and Mahmoud Payami$^{\ast}$}}\vspace{0.5cm}

\vspace{0.6cm}

\noindent \normalsize
{
Recently, some new series of heteroleptic ruthenium-based dyes, the so-called RD dyes, were designed and synthesized showing better 
performances compared to the well-known homoleptic N719. In this work, using the density-functional theory and its 
time-dependent extension, we have investigated the electronic structure and absorption spectra of these newly 
synthesized dyes, and compared the results to those of N3 dye to describe the variations of the properties due to 
the molecular engineering of ancillary ligand. We have shown that the calculation results of the absorption spectra 
for these dyes using the PBE0 for the exchange-correlation functional are in a better agreement with the experiment than using B3LYP or range-separated CAM-B3LYP. We have also derived a formula based on the DFT and used it to 
visually describe the level shifts in a solvent. The higher $J_{sc}$ observed in these new dyes is explained by the 
fact that here, in contrast to N3, the excitation charge was effectively transferred to the anchoring ligand. 
Furthermore, we have shown that the difference dipole moment vectors of the ground and 
excited states can be used to determine the charge-transfer direction in an excitation process.
Finally, the different electron lifetimes observed in these dyes is explained by investigating the adsorption geometries and the relative orientations of iodine molecules in different ``dye$\cdots$I$_2$'' complexes. 
}
\vspace{0.5cm}
 \end{@twocolumnfalse}
  ]

\section{Introduction}\label{sec1}


\footnotetext{\textit{Theoretical and Computational Physics Group, School of Physics and Accelerators, AEOI, 
P.~O.~Box~14395-836, Tehran, Iran; E-mail: mpayami@aeoi.org.ir}}

Solar technologies has experienced significant progress since the advent of dye-sensitized solar cells (DSSCs) in
 the early nineties.\cite{Oregan1991} DSSCs, as a low-cost alternative for traditional photovoltaic technologies,
have drawn the attention of research and industry communities over the past two decades. In a typical DSSC, the
sensitizers, which are adsorbed on TiO$_2$ nanoparticles, inject photo-excited electrons into the lower
unoccupied conduction band (CB) of the semiconductor. The injected electrons move through the load to the
counter-electrode and by reduction of I$_3^-$, regenerate I$^-$ ions in the electrolyte. Finally, the regenerated
ions reduce the oxidized dye molecules into their neutral states, and thereby, closing the 
circuit.\cite{hagfeldt2010}
The light harvesting photo-sensitizers play the most crucial role in the performance of a DSSC, and to enhance the 
efficiency, a large part of research activities were focused on the design and characterization of new 
sensitizers. 

A wide range of photo-sensitizers, including metal complexes,\cite{meyer2009} phthalocyanines,\cite{Martinez2010} 
zinc porphyrins,\cite{Yella2011,mathew2014} and metal-free organic dyes,\cite{Mishra2009} have been synthesized and 
used in 
DSSCs over the last years. Among the above-mentioned sensitizers, the ruthenium-based complexes have shown an 
impressive photovoltaic capabilities including broad absorption spectra, appropriate alignment of ground- and 
excited-state energy levels at the sensitizer/semiconductor interface, and a relatively good 
stability.\cite{hagfeldt2010} The homoleptic Ru complex, 
``$cis$-(SCN)$_2$bis(2,2$^{\prime}$-bipyridyl-4,4$^\prime$-dicarboxylic)ruthenium(II)'',
coded as N3, was the most famous one which played an important role in the improvement of DSC technology.  

Based on molecular engineering of N3, some new dyes were designed and synthesized aiming at: {\it i}) broadening 
the 
absorption spectra,\cite{nazeeruddin2001} {\it ii}) enhancing the light-harvesting capacity by either whole 
substitution of one of the bipyridine ligands which leads to heteroleptic families\cite{el2012novel,hussain2013structure,el2013influence,el2014structure,hussain2014comparative,cheema2014influence,cheema2014influence1,cheema2015femtosecond} or by introducing 
thiophene 
moieties,\cite{ANGE:ANGE200601463,Gao2008,Mishra2011} {\it iii}) increasing the chemical stability by replacing 
thiocyanate 
(SCN) 
ligand,\cite{bessho2009} and {\it iv}) reducing the recombination rate and increasing the 
dye-loading.\cite{Huang2010} In this respect, Huang and co-workers have designed and synthesized some new dyes 
which was based on replacing one of the 4,4$^\prime$-dicarboxylic-2,2$^\prime$-bipyridine (dcbpy) ligands in N3 
with a 
new benzimidazole (BI) contained one.\cite{Huang2010,Huang2012,Huang2013} The molecular structures of some of 
these dyes are compared with that of N3 in figure 1. In N3, each of the dcbpy equally can behave as an anchoring or 
ancillary ligands, wheras in RD dyes, the anchoring and ancillary roles are played separately by the dcbpy and 
BI-contained ligands, respectively.     

\begin{figure}
	\centering
		\includegraphics[width=0.75\columnwidth]{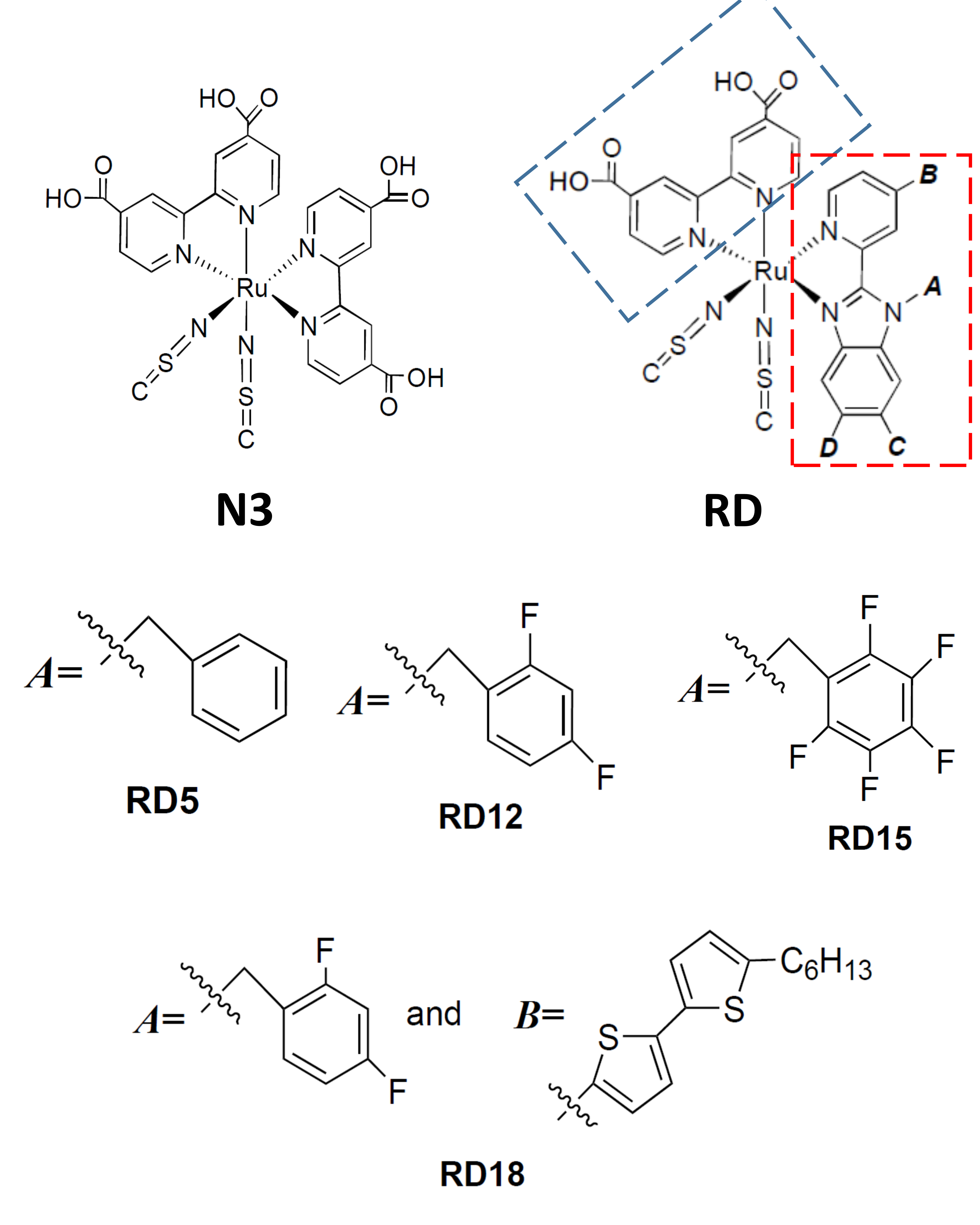}
		\caption{Molecular structures of N3 and RD dyes. Going from N3 to RD dyes, one of the dcbpy ligands (inside 
		blue rectangle) is kept unchanged, whereas the other one   
		has been replaced by a BI-contained ligand (inside red rectangle). Attaching of A and B 
		fragments at specified A and B positions result in RD5, RD12, RD15, RD18 structures.}
	\label{fig1}
\end{figure}

From historical point of view, RD dyes can be classified into three series. The best dye in the first 
series,\cite{Huang2010} RD5, showed a higher short-circuit current density ($J_{sc}$) compared to N719 (15.084 vs. 
14.157~mA/cm$^2$), but the cell performance was comparable to that of N719. 

Designing the second 
series\cite{Huang2012} was based on modification of the RD5 dye. The benzyl ring in RD5 was 
replaced by a fluorobenzyl ring with a varying number of fluorine atoms. In this engineering, while $J_{sc}$ 
decreases, the open-circuit voltage, $V_{oc}$, increases. The best achieved performance in this series, was for 
RD12 which contained two fluorine atoms. In this step, the performance of RD12 cell outpaced that of N719 cell 
(9.49 vs. 9.30). 
Although the performance of RD12 cell was higher than that of N719 cell, the extinction 
coefficients of both RD5 
and RD12 dyes were lower than that of N719. This observation motivated Diau's group to focus on the ways to 
increase the extinction coefficient. Knowing that, adding thiophene derivatives to pyridine part of the ancillary 
ligand could give rise to an enhancement in light harvesting capacity,\cite{ANGE:ANGE200601463} the third series 
were 
designed and synthesized.\cite{Huang2013} Among this set, RD18, containing two thiophene rings, turned out to 
be the optimum structure with $J_{sc}$ significantly higher but $V_{oc}$ slightly lower than those of N719. In that 
setup, the performance of RD18 cell was increased by 0.8\% compared to N719 cell (See Table 1 of  
ref.\citenum{Huang2013}).

In this work, we have employed density functional theory and its time-dependent extension (DFT\cite{HK64} 
and TDDFT\cite{RG84}) to study the electronic structure and absorption spectra of N3, RD5, RD12, RD15, and 
RD18 complexes, both in vacuum and in dimethylformamide (DMF) solvent. The calculation results for N3 are used here 
to describe the variations of the properties due to the structural modifications taken place in RD dyes. 

The energies of the frontier orbitals and the distribution of these orbitals over different ligands are 
calculated. The results show that for RD dyes, in contrast to N3, the distribution is not 
symmetric, and the HOMOs alternatively change the locations between two thiocyanate ligands whereas the LUMOs 
alternate between ancillary and anchoring ones. Moreover, we have shown that the energy shifts due to solvent are 
in the direction of a better alignment of HOMO and LUMO levels with the redox potential of the electrolyte and 
conduction band of the semiconductor, respectively. For a simple visual prediction of the direction of a level 
shift in a solvent, we have derived a formula based on DFT which is used in conjunction with the molecular 
electrostatic potential (MEP) plots
and orbital distributions over different atoms of a molecule. 
 
Analysing the excitation corresponding to the first peak of UV/vis spectra reveals that in RD dyes, in contrast to 
N3, the excited charge is transferred to the anchoring ligand, which in turn, enhances the effective charge 
injection to the nanoparticle. For a simple illustration of charge transfer direction in an excitation process, we 
have written a simple formula relating that direction to the difference dipole moment vectors of the ground and 
excited states.  

Finally, using the adsorption geometries and the orientations of iodine molecules in different ``dye$\cdots$I$_2$'' 
complexes, we have explored their interconnection with the different recombination rates observed in the RD dyes. 

The organization of paper is as follows. Section 2 is devoted to the computational details, the calculated results are presented and discussed in 
section 3, and we have concluded this work in section 4. 
\section{Computational details}\label{sec2}
We have determined the equilibrium geometries of neutral RD dyes within 
the B3LYP approximation\cite{B3LYP93-1,B3LYP93-2} for the exchange-correlation (XC) functional and 6-31+G(d) basis set in the DFT 
calculations using GAMESS-US package\cite{Schmidt1993} for both gas phase and in solvent. 

Because of its proper treatment of the polarization effects, we have used the polarized continuum 
model\cite{Cramer1999,Tomasi2005} (PCM) in which the solvent is assumed as a structureless dielectric medium and 
the solute is confined in a cavity which is formed from some overlapping spheres centred on 
atoms. In this calculations, we have used the most popular and fastest one of such models, called the 
conductor-like PCM\cite{Klamt93} (C-PCM). In the C-PCM, the surrounding medium is assumed as a conductor (with 
infinite dielectric constant), and the surface charge density is renormalized by a scaling function to result in an 
accurate charge density for the real medium with finite dielectric constant. 

The excitation energies and the oscillator strengths were calculated from solving the Casida 
equations\cite{Casida1995,Dreuw2005,Casida2009},
\begin{equation}\label{eq1}
  \begin{bmatrix}
   \textbf{A} & \textbf{B} \\
   \textbf{B}^* & \textbf{A}^* \\
  \end{bmatrix}
  \begin{bmatrix} 
\textbf{X}  \\
\textbf{Y}  \\
 \end{bmatrix}
 =
 \omega 
  \begin{bmatrix}
   \textbf{1} & \textbf{0} \\
   \textbf{0}& \textbf{-1} \\
  \end{bmatrix} 
  \begin{bmatrix} 
\textbf{X}  \\
\textbf{Y}  \\
 \end{bmatrix}
\end{equation}
To compare with experimental results, we have obtained the extinction coefficient from convolution of the 
calculated oscillator strengths by Gaussian functions with 
an appropriate FWHM, $\Delta_{1/2}$ as
\begin{equation}\label{eq2}
\epsilon(\omega)=2.174 
\times10^{8}\sum_I{\frac{f_I}{\Delta_{1/2}}exp[\;2.773\frac{(\omega_I^2-\omega^2)^2}{\Delta_{1/2}^2}\;]}
\end{equation}
where, $f_I$'s are the oscillator strengths, and $\omega_I$'s are the excitation frequencies. 

For the excited-state calculations in solvent, we have used non-equilibrium C-PCM/TDDFT in which it is assumed that 
the response of the solvent electrons to the "instantaneous" change of the solute charge distribution 
(due to the excitation) is very fast compared to that of the ions.\cite{Cossi2001,Marenich2011}
To calculate the vertical excitation energies, only the electronic response is considered and the solvent ions are 
assumed to be frozen at their locations.\cite{adamo2013}

Employing density-matrix based formulation of TDDFT,\cite{Furche2001} we have calculated the relaxed one-particle 
difference density matrix from which the first-order properties and partial charges in excited 
states\cite{Furche2002} are extracted.    

Full relaxed deposition geometry of the RD dyes on the surface of TiO$_2$ nanoparticles have been determined using 
both the periodic-slab and cluster methods. Using a $5\times 3$ monoclinic supercell along [010] and [11$\bar{1}$] 
directions, we have constructed an anatase 4-(TiO$_2$)-layer slab with (101) surface in the periodic-slab method.
The equilibrium geometries of the combined RD/slab and RD/cluster systems are calculated within the DFT and the 
self-consistent 
solution of the Kohn-Sham (KS) equations\cite{KS65} at the level of PBE generalized gradient approximation\cite{PBE96} employing the SIESTA 3.2 code package and using a split-valence 
double-$\zeta$ basis set augmented by polarization functions (DZP) along with the existent nonrelativistic pseudopotentials for Ti, O, C, N, S, F, H, and Ru atoms. The cutoff for the 
plane-wave was chosen as 200 Ry to assure the conformation of our results with those obtained using the Quantum 
ESPRESSO code package.\cite{QE-2009}  
For the cluster calculations, an anatase (TiO$_2)_{38}$ cluster is used to model the 
nanoparticles.\cite{Persson2000,Pastore2012chem,deangelis2008}

Geometries of  ``RD$\cdots$I$_2$'' and ``bithiophene$\cdots$I$_2$'' complexes were optimized using NWChem code 
package\cite{Valiev2010} with the highly polarized 6-311G(d,p) basis set within B3LYP approximation. The interaction 
energies between I$_2$ and RD/bithiophene were calculated using
\begin{equation} \label{eq3}
E_{\rm int}=E_{\;\rm X\cdots I_2}-(E_{\;\rm I_2}+E_{\;\rm X})-\Delta E_{\;\rm CP}    
\end{equation}
in which $E_{\;{\rm X}\cdots {\rm I}_2}$ is the total energy
 of the ``RD$\cdots$I$_2$'' or ``bithiophene$\cdots$I$_2$'' 
complexes, $E_{\;\rm I_2}$ and $E_{\;\rm X}$ are the total energies of the isolated components, and $\Delta 
E_{\;\rm CP}$ stands for the compensation correction for the basis set superposition error (BSSE).\cite{boys1970}

For visualization of structures, densities, and molecular orbitals we have used VESTA\cite{momma2008vesta}, MacMolPlt\cite{bode1998macmolplt}, and VMD\cite{humphrey1996vmd} graphical interfaces.
\section{Results and discussion}\label{sec3}
\subsection{Equilibrium properties of RD dyes}
Geometrical structures for RD dyes have been fully optimized using GAMESS-US, in both vacuum and DMF at the 
B3LYP/6-31+G(d) level of theory, and some selected geometrical parameters for N3, RD5, and RD12 are summarized in 
Table~\ref{tab1}. 

\begin{table}
\caption{Selected bond lengths (in $\AA$) and  angles (in degrees) for N3, RD5, and RD12 dyes in gaseous phase 
 and in DMF.}
 \centering
\resizebox{\columnwidth}{!}
{
\begin{tabular}{@{}lcccccccc} \hline \hline 
Parameters		& \multicolumn{2}{l}{N3}& & \multicolumn{2}{l}{RD5}  & &  \multicolumn{2}{l}{RD12}  \\ \cline{2-3} 
\cline{5-6}  \cline{8-9} 
                        & Gas    & DMF      &         & Gas     & DMF      & & Gas   & DMF  \\  \hline
d$_{Ru-N1}$	  		    & 2.08   &  2.09	&         & 2.11	& 2.12    & &  2.11 & 2.12  \\
d$_{Ru-N2}$				& 2.09   &  2.08	&         & 2.11	& 2.10    & &  2.11 & 2.10   \\
d$_{Ru-N3}$				& 2.08   &  2.09	&         & 2.06	& 2.08    & &  2.06 & 2.08   \\
d$_{Ru-N4}$				& 2.08   &  2.08	&         & 2.06	& 2.07    & &  2.06 & 2.07   \\
d$_{Ru-N5}$				& 2.06   &  2.08	&         & 2.05	& 2.08    & &  2.05 & 2.08	  \\
d$_{Ru-N6}$				& 2.06   &  2.08	&         & 2.09	& 2.09    & &  2.07 & 2.08	  \\
$\theta_{N1-Ru-N2}$     &  78.5  &  78.5	&         & 77.1	& 77.0    & &  77.0 & 76.9	  \\
$\theta_{N2-Ru-N3}$     & 99.6   &  98.4	&         & 102     & 101     & &  102  & 101   \\
$\theta_{N3-Ru-N4}$     & 78.5   &  78.5	&         & 78.9	& 78.7    & &  78.9 & 78.7	  \\
$\theta_{N2-Ru-N4}$     & 94.3   &  92.4	&         & 95.1    & 92.3    & &  94.3 & 92.1	  \\
$\theta_{N5-Ru-N6}$     & 92.5   &  90.1	&         & 92.8    & 90.0    & &  92.5 & 90.3    \\  
$\theta_{Ru-N5-S}$      &  171   &  179	    &         & 176     & 179     & &  179  & 179    \\
$\theta_{Ru-N6-S}$      &  170   &  179	    &         & 152     & 177     & &  159  & 176     \\    
\hline                                                          
\end{tabular}
}
\label{tab1}
\end{table}

Examining the values listed in Table~\ref{tab1}, shows that in both phases, the Ru-N bonds and the angles between 
them have almost the same values for the N3, RD5, and RD12 dyes. Moreover, the Ru-N bond lengths in the solvent (in 
PCM framework) are slightly greater than those in gas phase.
\begin{figure}
 	\centering
 		\includegraphics[width=0.5\columnwidth]{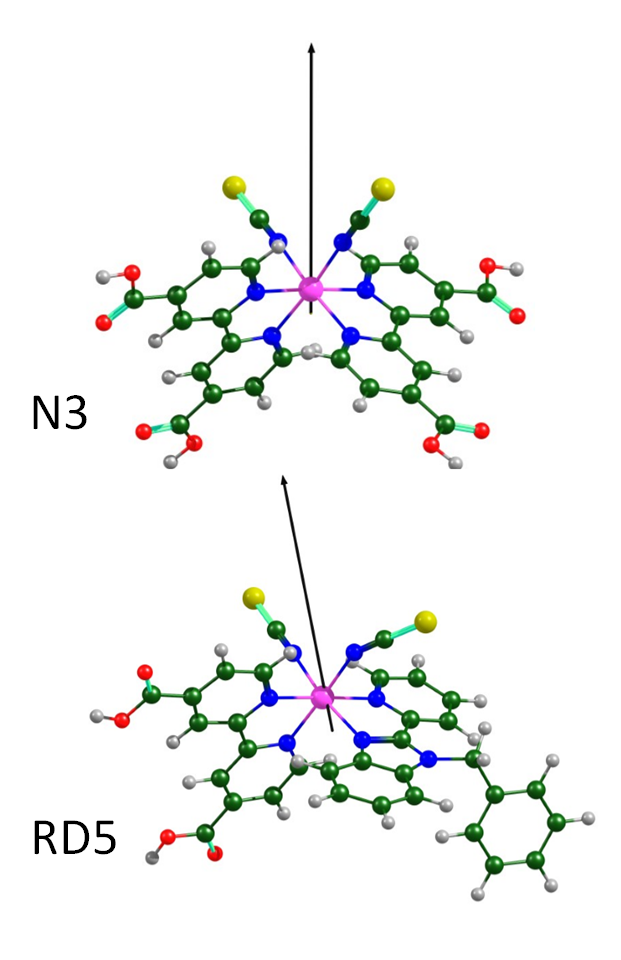}
 		\caption{The molecular structures of N3 and RD5 in vacuum, showing the bending of Ru-thiocyanate bond in 
 		RD5. The arrows indicate the directions of electric dipole moments. }
 	\label{fig2}
 \end{figure}
On the other hand, in all cases, the weakening of the 
interaction between ligands in the solvent,\cite{Chen2007} almost removes any deviations from an ideal octahedral 
structure.
The values in the last two rows of Table~\ref{tab1} show a decrease in bending of Ru-thiocyanate bond upon going from gas phase to the solvent (see Fig.~\ref{fig2}), which is due to the weakening of the S-$\pi$ 
interactions between sulphur and BI group.\cite{Tauer2005} Disappearing of the bending in the solvent 
implies that the S-$\pi$ interactions have an electrostatic character.
\subsection{Electronic structure}
\begin{figure}
 \centering
 	\includegraphics[width=\columnwidth]{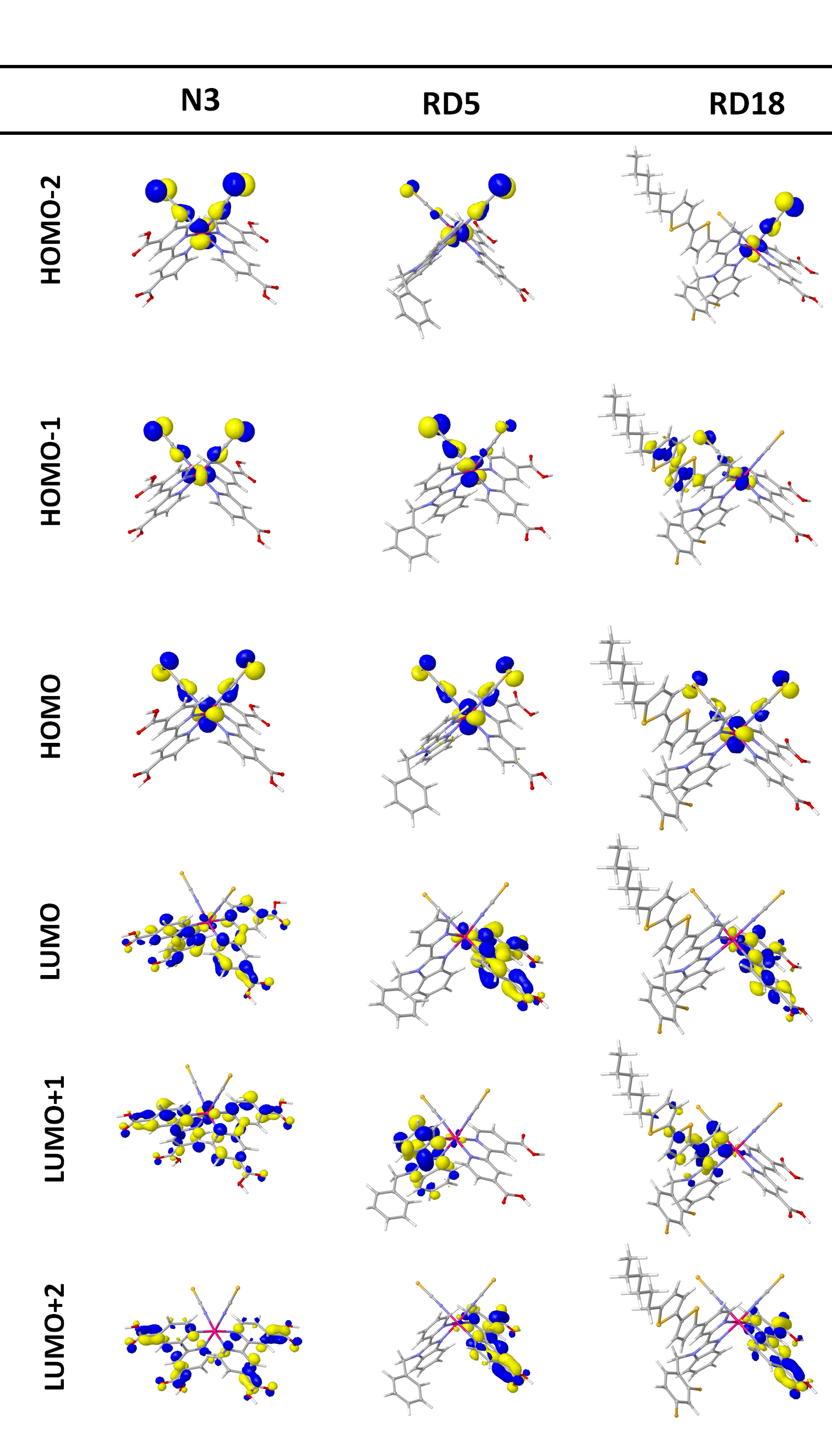}
 		\caption{Frontier molecular orbitals of N3, RD5, and RD18 dyes. The HOMOs and LUMOs of RD dyes are 
 		assymetrically distributed over the ligands while in N3 the distribution is symmetric. In RD18, the HOMO-1 
 		and LUMO+1 are partly localized on the thiophene rings.}
 	\label{fig3}
 \end{figure}

The spatial distribution of the frontier molecular orbitals in a dye molecule plays a significant role in the 
effective charge injection from that molecule to the semiconductor nanoparticle. To understand the effects of 
substitution of BI ligands, we have plotted the isodensities of the frontier molecular orbitals for N3, RD5, and 
RD18 dyes in figure~\ref{fig3}.
  
As shown in figure~\ref{fig3}, the HOMOs in both N3 and RD5 dyes
are distributed over the central Ru and two SCN ligands, while in RD18 the thiophene rings have 
tangible contributions in the HOMO-1. 
The LUMOs in N3 are distributed over dcbpy ligands whereas in RD dyes, they 
are distributed over dcbpy and BI-contained ancillary ligands. 
For N3 dye, because of its symmetric geometry, the distribution of HOMOs over SCN  and LUMOs over 
anchoring-ancillary ligands are symmetric. However, in RD dyes, because of BI-substitution, the geometry is not 
symmetric anymore and consequently, the distribution symmetry is spoiled for RD dyes such that, for example, the 
LUMO and LUMO+1 are localized on the anchoring and
ancillary ligands, respectively. 
As we will show in the following subsection, the LUMO and LUMO+1 have significant 
contributions in the optical transitions corresponding to the first absorption peak. 

\begin{figure}
 	\centering
 		\includegraphics[width=0.9\columnwidth]{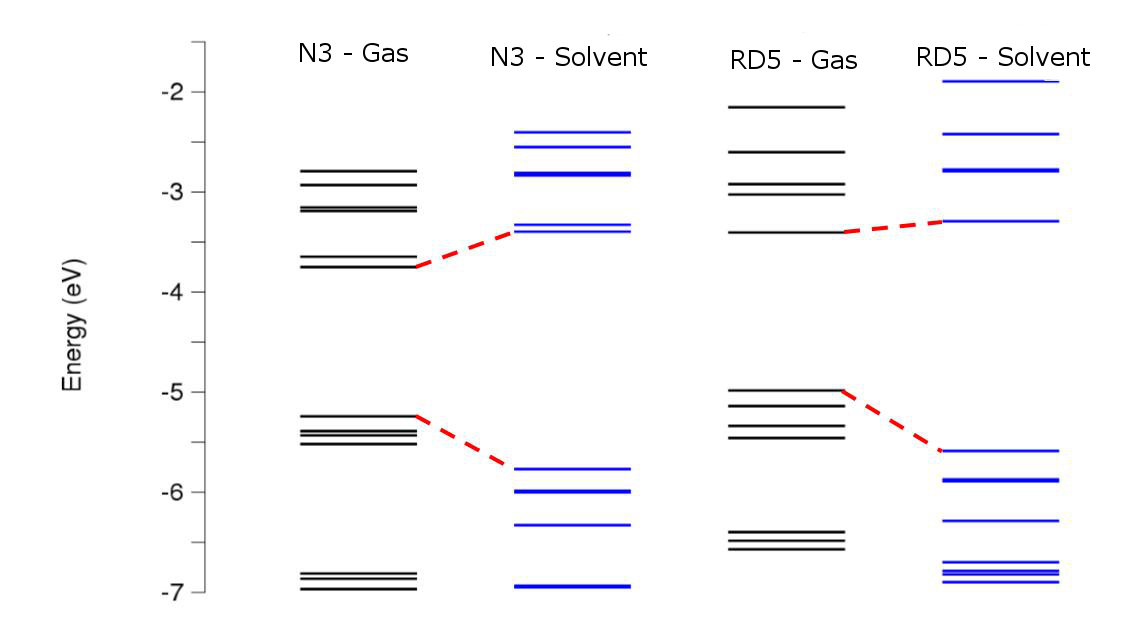}
 		\caption{Energy levels (in eV) of frontier orbitals of N3 and RD5 both in vacuum and solvent. The HOMO-LUMO 
 		gaps are increased in the solvent for both dyes.}
 	\label{fig4}
 \end{figure}

To explore the solvent effects on the electronic structure of these dye molecules, we have plotted, in 
figure~\ref{fig4}, their corresponding energy levels of the frontier molecular orbitals, both in vacuum and  
solvent. 
As is shown in the figure, for both dyes, all LUMOs have been destabilized while all HOMOs have  
been stabilized in the solvent. The shifts in the levels lead to the widening of the HOMO-LUMO gaps. 
The calculation results for N3 (RD5) molecule show a destabilization by 0.3 (0.1)~eV for the LUMO and a stabilization by 0.5~eV for the 
HOMO levels. Therefore, the gap widening is 0.8 (0.6)~eV for N3 (RD5). In the following, we have 
given a simple formulation for the amount and direction of 
the level shifts and have used to describe the shifts of the frontier orbitals. 

\subsubsection{DFT formulation of level shifts in solvent}\hspace{1mm}

When a molecule is inserted in a cavity surrounded by a dielectric medium, the charge density of the molecule 
polarizes the dielectric, and the local polarization creates a local electric field which, in turn, interacts with 
the initial charge distribution. The interaction modifies the initial charge distribution, and consequently, the 
dielectric polarization is modified. This cycle continues until the charge density does not change any more. Here 
we consider only the first cycle which gives the leading correction term.
Introducing of a local electric field, $\EE(\rr)$, to a many-electron system, the correction in the electronic part 
of the Hamiltonian is given by
\begin{equation}\label{eq4}
\Delta \hat{V}=+e\sum_{i=1}^{N} {\EE}({\rr_i})\cdot{\rr_i}
\end{equation} 
where, the centre of positive charges is chosen as the origin of the coordinate system.
Using the language of the DFT, the correction in the total energy functional appears as
\begin{equation}\label{eq5}
\Delta W = +e\int \rho(\rr)\;\rr\cdot\EE(\rr)\;d\rr 
\end{equation}
The above correction term in the total energy, contributes the correction term
\begin{equation}\label{eq6}
\Delta v^{\rm KS}(\rr)=+e\rr\cdot\EE(\rr)
\end{equation}
in the KS equations. Using the recipe of the first-order energy correction in the perturbation theory, the level 
shifts are given by
\begin{eqnarray}\label{eq7}
\Delta \varepsilon_i&=&\langle\psi_i^{(0)}(\rr)\mid\ +e\rr\cdot\EE(\rr)\mid\psi_i^{(0)}(\rr)\rangle \nonumber \\ 
                    &=&+e\int \rho_i^{(0)}(\rr)\rr\cdot\EE(\rr)\;d\rr 
\end{eqnarray} 
where $\psi_i^{(0)}$ is the KS orbital calculated in vacuum.
Now, if the cavity is composed of some branches specified by the set of position vectors $\{\RR_\alpha\}$, and 
assuming the effective electric field is constant and having a proper direction inside a branch, then equation~(\ref{eq7}) reduces approximately 
to
\begin{eqnarray}\label{eq8}
\Delta \varepsilon_i&\approx&\sum_\alpha \left[ +e\int 
\rho_{i,\alpha}^{(0)}(\rr)\rr\;d\rr\right]\cdot\EE(\RR_\alpha)\nonumber \\
                    &=&-\sum_{\alpha} {\bm \mu}_{i,\alpha}^{(0)}\cdot\EE(\RR_\alpha) 
\end{eqnarray}
where, 
$\rho_{i,\alpha}^{(0)}(\rr)$ and ${\bm \mu}_{i,\alpha}^{(0)}\;(\equiv -e\int \rho_{i,\alpha}^{(0)}(\rr)\rr\;d\rr)$
are the contributions (to the total charge density, $\rho_i^{(0)}(\rr)$, and dipole moment, ${\bm \mu_i}$, of the 
orbital $\psi_i^{(0)}$, respectively) of that part of KS molecular orbital that is localized inside the $\alpha$ branch of the 
cavity, satisfying
\begin{equation}\label{eq9}
\sum_\alpha {\bm \mu}_{i,\alpha}^{(0)}={\bm \mu_i}^{(0)}\;\;\; {\rm and}\;\;\; \sum_\alpha 
\rho_{i,\alpha}^{(0)}(\rr)=\rho_i^{(0)}(\rr)
\end{equation}

\begin{figure}
 \centering
 	\includegraphics[width=.8\columnwidth]{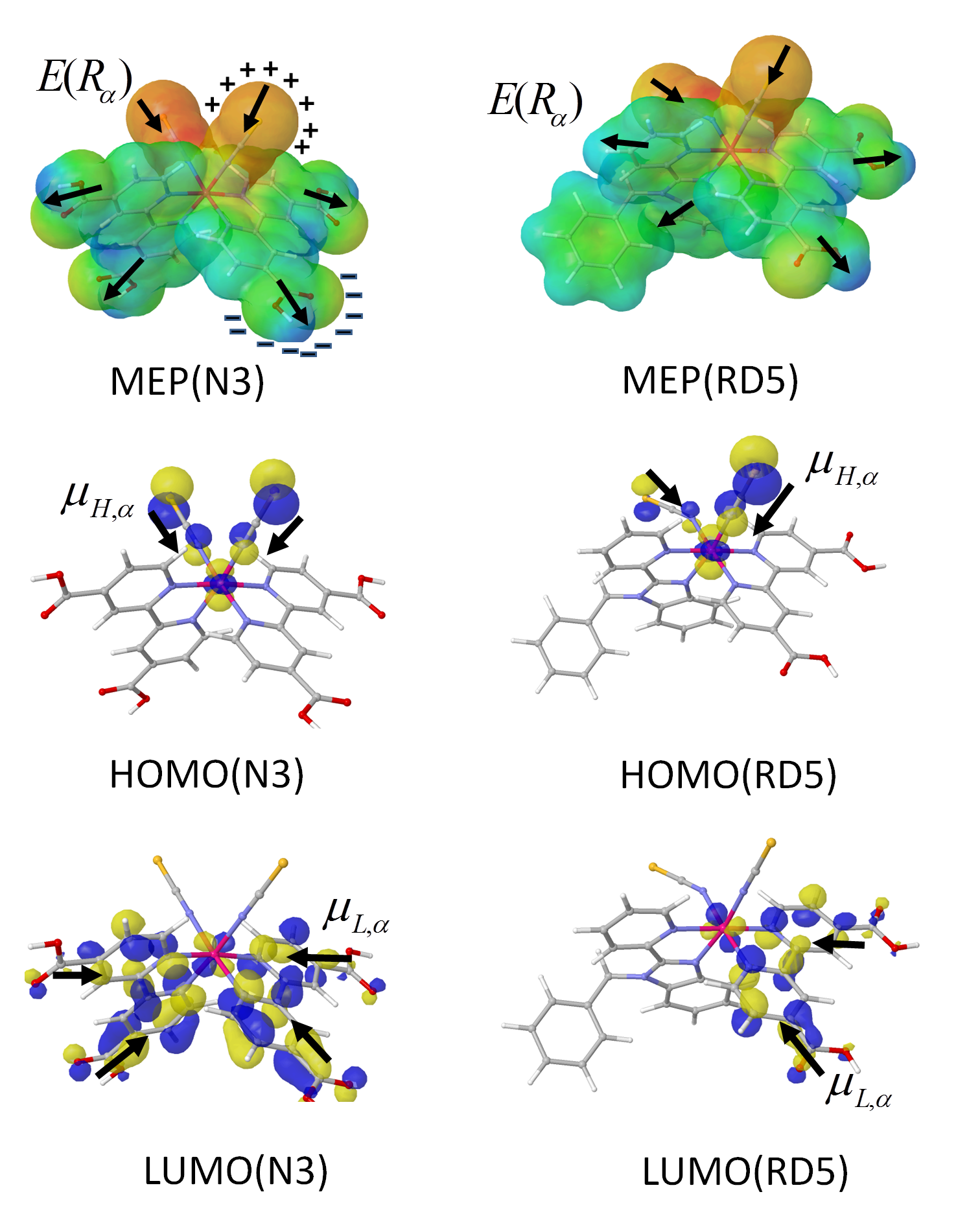}
 		\caption{MEP plot as well as the HOMO and LUMO densities for N3 and RD5 
 		in vacuum. The arrows in MEP plot indicate local electric fields inside the cavity branches, while the arrows in the HOMO and LUMO plots correspond to electric dipole 
 		moments inside the cavity branches of those orbitals.}
 	\label{fig5}
 \end{figure}

To explain the level shifts in the solvent, we have plotted, in figure~\ref{fig5}, the MEP maps as well as the HOMO 
and LUMO densities for N3 and RD5 molecules. 

The cavity boundaries in the PCM resembles the density isosurface in the MEP plot, the reddish and bluish colours 
of the plot specify the electron-rich and electron-deficient regions, respectively. The electron-rich and 
electron-deficient parts of the molecule induce positive and negative charges, respectively, on the cavity surface 
which give rise to local electric fields inside the cavity. 

As shown in figure~\ref{fig5}, for HOMOs the dipole 
moments and electric fields are ``in the same direction'', while for LUMOs they are ``in opposite directions''.
The HOMOs for N3 and RD5 are distributed over two branches and equation~(\ref{eq8}) predicts more or less the same 
shifts (stabilizing) in good agreement with those shown in figure~\ref{fig4}. The LUMO of N3 is distributed over 
four branches whereas that of RD5 is distributed over two branches (half of that for N3). Since in the LUMO case, 
the directions of the dipoles and fields are opposite, the shifts are toward destabilization (upward) and the 
magnitude for N3 is about two times that for RD5, in excellent agreement with results shown in figure~\ref{fig4}.

\subsubsection{Population analysis}\hspace{1mm}
To carry out the population analysis, we consider each ruthenium complex as consisting of three different parts 
(see figure~\ref{fig1}): 
``Ru(SCN)$_2$'', ``anchoring'', and ``ancillary'' ligands (for N3, the ancillary and anchoring ligands are 
equivalent). The RD dyes are formed by substituting one of the two equivalent ligands in N3 by a BI-contained ancillary 
ligand. Using the analysis results for the three parts of the complex in its ground state, we determine the amount 
of charge migration resulted from each substitution, and comparing them with those of the excited state gives us 
the direction of charge transfer in an excitation process. The calculated L\"owdin partial charges for the three 
parts, both in ground and excited states are listed in Table\ref{tab2}.  
\begin{table}
   \caption{L\"owdin partial charges (in atomic units) for three specified parts of RD dyes for the ground and 
   excited states. For excited states, (except for RD18 which is S$_3$) only S$_5$ excitations contribute to the 
   first peak. The magnitudes of dipole moments (in Debye) are listed in the last column.}
\begin{center}
\resizebox{0.9\columnwidth}{!}
{
   {\begin{tabular}{@{}llcccc} \hline \hline 
   &    &\multicolumn{3}{c}{L\"owdin charge (e) }             &  $\mu$ (Debye)    \\  \cline{3-5} 
   &                         &  Ru(SCN)$_2$  &  dcbpy      & ancillary   &              \\  \hline
  \multirow{1}{*}{N3}   & GS &   -1.500      &  +0.750     &   +0.750    &   22.42      \\
                        &  ES&   -0.947      &  +0.473     &   +0.474    &   11.59      \\
               & ES-GS       &   +0.553      &  -0.277     &   -0.276    &             \\
&&&&& \\
  \multirow{1}{*}{RD5} &  GS &   -1.546      &  +0.709     &   +0.837    &  27.22      \\
                    &     ES &   -1.039      &  +0.273     &   +0.766    &  20.39      \\
               &  ES-GS      &   +0.507      &  -0.436     &   -0.071    &             \\
&&&&& \\
  \multirow{1}{*}{RD12} &  GS&   -1.528      &  +0.706     &   +0.822    &  26.50     \\
                &  ES        &   -1.018      &  +0.293     &   +0.725    & 18.79            \\
                 &  ES-GS    &   +0.510      &  -0.413     &   -0.097    &             \\
&&&&& \\
  \multirow{1}{*}{RD15} &  GS&   -1.530      &  +0.698     &   +0.832    &  25.89      \\
             &     ES        &   -1.017      &  +0.355     &   +0.662    &  17.81           \\
                 &  ES-GS    &   +0.513      &  -0.343     &   -0.170    &             \\
&&&&& \\
  \multirow{1}{*}{RD18} &  GS&   -1.495      &  +0.693     &   +0.802    &  30.20      \\
             &            ES &   -1.004      &  +0.673     &   +0.331    &  18.49           \\
                 &  ES-GS    &  +0.491       &  -0.019     &   -0.472    &             \\   \hline
       \end{tabular}}
 }      
\end{center} 
    \label{tab2}
  \end{table}

The calculation results for the ground state show that, the geometric symmetry in N3 leads to an equal distribution of positive charges 
over the two dcbpy ligands and an equal negative charges over the two thiocyanate ligands. Direct calculation of dipole moment from electronic charge density shows that the vector lies on the bisector of the angle formed by the two SCN groups (see figure~\ref{fig2}) consistent with the charge symmetry from population analysis.
    
For RD dyes, the amount of positive charge on the BI-contained ligand (ancillary) is more than that of the 
corresponding ligand on N3 (dcbpy), whereas the thiocyanate and anchoring ligands are less positive compared to the 
corresponding ligands in N3 which implies an electron migration from ancillary to other parts. The calculation of 
dipole moments for these dyes show that the dipole vectors do not coincide with the angle bisector any more (see 
figure~\ref{fig2}) which is consistent with the electron migration found in population analysis. The analysis results for excited states will be used in the discussion of absorption spectra. 

\subsection{Absorption spectra}
 In the context of the TDDFT, we have calculated the excitation energies and oscillator strengths of the lowest 60 
 excitations for N3 and RD dyes. Taking more excitations into account, did not affect the absorption spectra in the visible region. 
The RD dyes are observed to have two different stereoisomeric structures with equal relative 
abundance, having no sensible differences in the band structure and optical properties.\cite{Huang2012,Huang2013}
We have therefore, considered only one of the isomers (the so-called ``A-isomer'', as in ref.~\citenum{Huang2012}) in 
our optical properties' 
calculations. 

\begin{figure}
 \centering
 	\includegraphics[width=0.9\columnwidth]{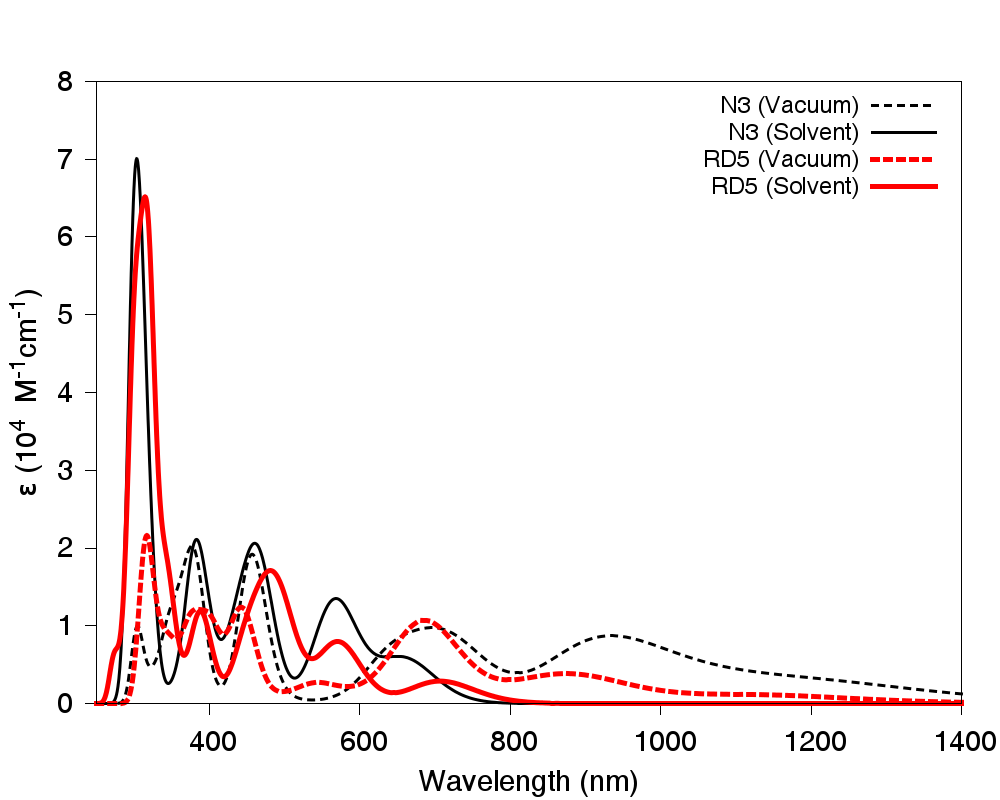}
 		\caption{Absorption spectra of N3 and RD5 in vacuum (dashed lines) and solvent (solid lines). The blue 
 		shift due to solvent is evident in both dyes.}
 	\label{fig6}
 \end{figure}

The solvatochromic effects were investigated by performing calculations for the absorption spectra of N3 and 
RD5 dyes in 
both vacuum and solvent, with the result shown in figure~\ref{fig6}.
As figure~\ref{fig6} shows, the solvent gives rise to blue shifts in the first peak of absorption spectra for N3 (0.53~eV) and RD5 (0.35~eV) dyes, consistent with the widening of HOMO-LUMO gaps shown in figure~\ref{fig4}. These blue shifts may also be explained by using the fact that the dipole moments in the excited states are smaller than their corresponding ground state vectors.\cite{azar2014} 

\begin{figure}
 \centering
 	\includegraphics[width=\columnwidth]{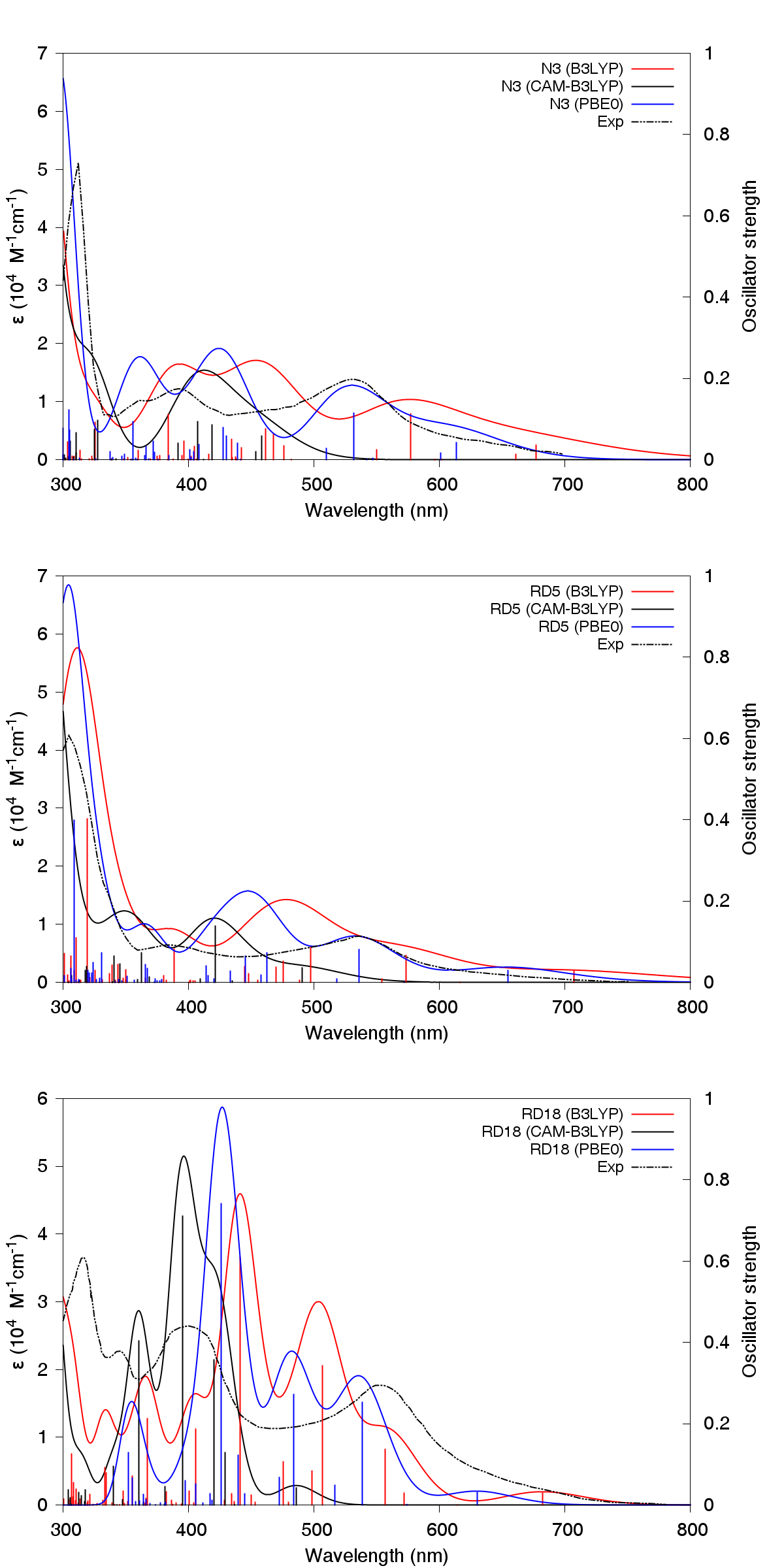}
 		\caption{Absorption spectra, obtained using different XC functionals, are compared with experimental
 		 results. The top, middle, and bottom subfigures correspond to N3, RD5, and RD18, respectively. Red, black, 
 		 blue solid lines correspond to B3LYP, CAM-B3LYP, and PBE0, respectively; the experimental results are 
 		 shown by dotted-dashed lines. The experimental data for N3 is from ref.~\citenum{yin2009enhanced}, and 
 		 those for RD5 and RD18 are from ref.~\citenum{Huang2012} and ref.~\citenum{Huang2013}, respectively.  The 
 		 bars represent the positions and values of oscillator strengths.}
 	\label{fig7}
 \end{figure}

Recent studies on excitation energies have shown that for charge-transfer (CT) excitations, the TDDFT calculations 
may lead to errors of the order of some eV,\cite{Dreuw2004,liao2003performance,perpete2006ab} and the use of 
range-separated XC functionals has therefore been prescribed.\cite{Peach2008} The diagnostic parameter $\Lambda$, 
which quantifies the charge-transfer character of excitations\cite{Peach2008,azar2014} and takes the values $0\le\Lambda\le 1$, are calculated for the 
dominant transitions of RD dyes and the results are listed in Table~\ref{tab3}. The calculated $\Lambda$ values are based on the B3LYP approximation. Small and large values of $\Lambda$ correspond to the CT and local character of excitations, respectively. The small values of $\Lambda$ listed in Table~\ref{tab3} indicates that all of the dominant transitions have CT characteristics, which originates from the small overlap integrals of frontier occupied with unoccupied orbitals (See figure~\ref{fig3}).  
The calculated absorption spectra using range-separated CAM-B3LYP\cite{Yanai2004} as well as PBE0\cite{adamo1999toward} 
approximations are compared with B3LYP and experimental results in figure~\ref{fig7} for N3, RD5, and RD18.
As is seen from figure~\ref{fig7}, PBE0 gives the best agreement with experiment whereas the CAM-B3LYP results 
are significantly blue-shifted. A similar behaviour has already been reported for a set of various ruthenium based 
complexes.\cite{bahers2014} This blue shift of CAM-B3LYP results can be attributed to the overestimation\cite{few2015models} of electron-hole binding energies for the ruthenium based complexes.    

\begin{figure}
 \centering
 	\includegraphics[width=\columnwidth]{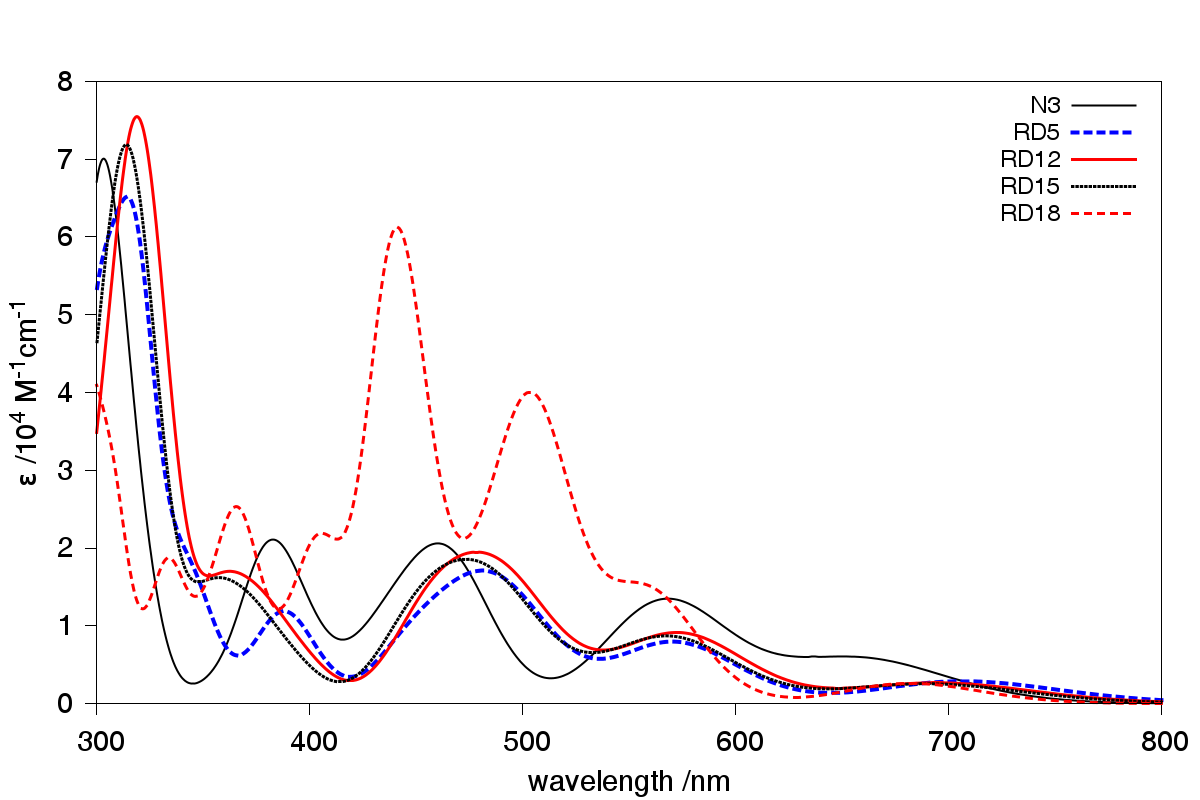}
 		\caption{Absorption spectra of N3 (solid black), RD5 (dashed blue), RD12 (solid red), RD15 (dashed green), 
 		and RD18 (dashed red) in DMF solvent. For RD18, the significant hyperchromic 
 		shifts result from attaching bithiophene unit to the BI-contained ligand.}
 	\label{fig8}
 \end{figure}

In order to get insight into the differences between N3 and the RD dyes, the absorption spectra of N3 and RD dyes 
are compared in figure~\ref{fig8}. Here the comparison is for the B3LYP calculation results which is sufficient for 
our purposes. 
This comparison for RD5, RD12, and RD15 reveals that the variation of the number of fluorine atoms does not affect the spectra. However, 
the result for 
RD18 shows that attaching a bithiophene unit to the BI-contained ligand gives rise to a significant enhancement in the extinction coefficient. 

\begin{table}
\caption{Excitation energies (EE), Oscillator strengths (f$_I$), diagnostic parameter $\Lambda$, and the excitation characters for different excited states (ES) around the first absorption peak. Those with f$_I < 0.01$ are not included.  } 
 \centering
 \resizebox{0.9\columnwidth}{!}
 {
  \begin{tabular}{llllll}
 	\hline
 Dye	 & ES    &EE(eV)& f$_I$ & $\Lambda$ & Character                                  \\    \hline
 	 &       &      &       &                                            \\
 RD5	 & S$_1$ & 1.75 & 0.027&0.35 & H\quad\:$\rightarrow$ L\quad\:(0.904)      \\ 
 	 & S$_3$ & 2.16 & 0.066 &0.34& H-2     $\rightarrow$ L\quad\:(0.785)           \\ 
 	 & S$_5$ & 2.49 & 0.087 &0.29& H-2     $\rightarrow$ L\quad\:(0.683)      \\ 
 	 &       &      &      & & H-1     $\rightarrow$ L+1(0.202)           \\  
 	 &       &      &      & &                                            \\
 RD12 & S$_1$ & 1.75 & 0.026&0.34 & H\quad\:$\rightarrow$ L\quad\:(0.905)      \\ 
 	 & S$_3$ & 2.16 & 0.065 &0.34& H-2     $\rightarrow$ L\quad\:(0.777)      \\ 
 	 & S$_5$ & 2.49 & 0.087 &0.28& H\quad\:$\rightarrow$ L+2(0.702)           \\ 
 	 &       &      &      & & H-1     $\rightarrow$ L+1(0.202)           \\ 
 	 &       &      &      & &                                            \\
 RD15 & S$_1$ & 1.75 & 0.025 &0.34& H\quad\:$\rightarrow$ L\quad\:(0.912)      \\ 
      & S$_3$ & 2.15 & 0.065 &0.33& H-2     $\rightarrow$ L\quad\:(0.721)      \\ 
      &       &      &      & & H\quad\: $\rightarrow$ L+1(0.121)           \\ 
      & S$_5$ & 2.48 & 0.087 &0.28& H\quad\:$\rightarrow$ L+2(0.592)           \\ 
      &       &      &      & & H-1     $\rightarrow$ L+1(0.306)           \\ 
 	 &       &      &       & &                                            \\
 RD18 & S$_1$ & 1.76 & 0.028 &0.35& H\quad\:$\rightarrow$ L\quad\:(0.907)      \\ 
      & S$_3$ & 2.07 & 0.184 &0.27& H\quad\:$\rightarrow$ L+1(0.640)           \\ 
      &       &      &       & & H-2     $\rightarrow$ L\quad\:(0.214)      \\ 
      & S$_5$ & 2.31 & 0.273 &0.35& H\quad\:$\rightarrow$ L+1(0.612)           \\ 
      &       &      &      & & H\quad\:$\rightarrow$ L+2(0.142)           \\ 
      &       &      &      & & H-1     $\rightarrow$ L+2(0.137)           \\   \hline
 	 
 	 \end{tabular} 
 }
 \label{tab3}
 \end{table}

To determine the excitation characters in the region around the first peak, the single-particle contributions for 
the first (S$_1$), third (S$_3$) and fifth (S$_5$) excitations are listed in Table~\ref{tab3}. 
The second and fourth ones, because of 
their negligible oscillator strengths, were not included. According to the results, the H$\rightarrow$L 
(i.~e., HOMO$\rightarrow$LUMO) transition has 
the dominant contribution ($\sim$90\%) in the first excitation (S$_1$) for all dyes. The distributions of frontier 
orbitals, which was discussed earlier, show that the S$_1$ excitation is accompanied by a charge transfer 
from the Ru atom and SCN ligands to the dcbpy anchoring ligand, in all RD dyes. On the other hand, in S$_3$ 
excitations, the H$\rightarrow$(L+1) is 
dominant for RD18, while for the other RD dyes the (H-2)$\rightarrow$L transition dominates. Accordingly, the 
charge transfer in 
RD18 is from the two (SCN) donor units to the ancillary ligand, whereas for other RD dyes, as in S$_1$, it is from 
the Ru atom and SCN ligands to the dcbpy anchoring ligand. Finally, as to the S$_5$ excitation (having 
the largest oscillator strength), which plays the dominant role in the build-up of the first absorption peak,
the two H$\rightarrow$(L+2) and (H-1)$\rightarrow$(L+1) transitions 
have significant contributions, and by going from RD5 to RD15 (which is accompanied by increasing the number of 
fluorine atoms), the weight of former changes from 68\% to 59\%, while that of the latter increases from 20\% to 30\%. Taking into 
account the distribution of the frontier orbitals, this increase of 
the second contribution (decrease of the first contribution) can be attributed to the decrease of the 
transferred charge to the anchoring ligand, which shows up as decrease in $J_{sc}$ 
in experimental results.\cite{Huang2012} 
For RD18, the contributions from H$\rightarrow$(L+1), 
H$\rightarrow$(L+2), and (H-1)$\rightarrow$(L+2) transitions are significant with values of $61\%$, $14\%$, and 
$13\%$, respectively. The largest contribution in these excitations corresponds to the charge transfer from the 
donor unit to the ancillary ligand. 

The values of transferred charge in S$_5$ excitations, listed in Table~\ref{tab2}, imply that 
for heteroleptic dyes, the amount of transferred charge to the ancillary and anchoring ligands are different, in 
contrast to the case of N3 dye in which it is the same for both ligands. On the other hand, with increasing the fluorine 
atoms, the amount of transferred charge to the anchoring ligand decreases, which is apparently due to the high 
electronegativity of fluorine atom. Since the extinction coefficients of RD5, RD12, and RD15 are more or less the 
same at all wavelengths in the visible region (Fig.~\ref{fig7}), the higher amount of charge transfer implies the 
higher value of the $J_{sc}$, in agreement with the observed experimental values.\cite{Huang2012} 

\begin{figure}
 \centering
 	\includegraphics[width=0.9\columnwidth]{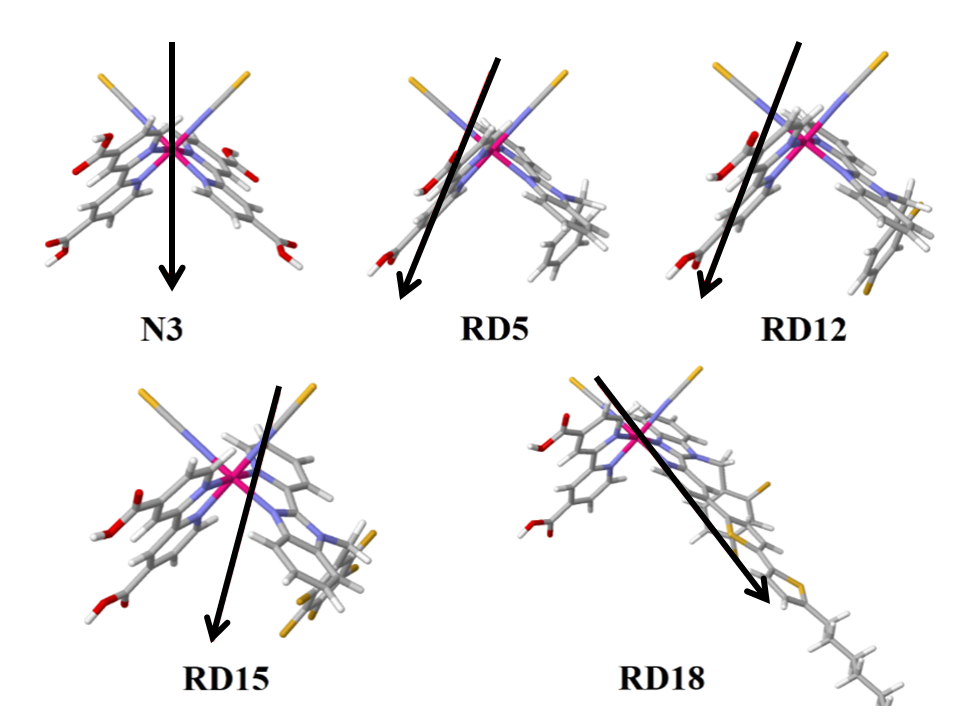}
 		\caption{The difference of the excited- and ground-state dipole moments shown by black solid arrows, 
 		determine the direction of charge transfer in the course of excitation.}
 	\label{fig9}
 \end{figure}

In a vertical excitation, since the ions do not change their positions, the difference of the excited- and 
ground-state dipole moments is related by: 
 \begin{eqnarray}\label{eq10}
({\bf r}_{ES}^-  -{\bf r}_{GS}^-)Q_t&=&[({\bf r}_{ES}^-  -{\bf r}_{ES}^+)- ({\bf r}_{GS}^-  -{\bf r}_{GS}^+)]Q_t \nonumber\\
        & =& {\bm \mu}_{ES}-{\bm \mu}_{GS}
 \end{eqnarray}
 to the difference in the centre-of-charge vectors (i.~e., $\Delta {\bf r}^-\equiv {\bf r}_{ES}^-  -{\bf r}_{GS}^-$), and can be used to determine the direction of charge transfer in the course of excitation (see figure~\ref{fig9}). This method is free from the ambiguities of net charge assignments to the atoms, that arise using different population analysis methods. 

In N3, because the charge transfer is towards the anchoring ligands in an equal footing, the difference vector is
the bisector of the angle formed by two SCN ligands.
However, for RD5, RD12, and RD15, the vector is oriented towards the anchoring than the ancillary ligand which 
implies that the larger fraction of charge is transferred to the anchoring ligand. In RD18, because of its 
thiophene rings, the vector is oriented towards the ancillary ligand.     

\subsection{Adsorption geometry of RD dyes}
Because of the carboxylic anchoring groups on both bipyridine ligands, there are many possible 
adsorption configurations for the homoleptic Ru-complexes. These chromophores could attach to TiO$_2$ surface 
through two or three carboxylic anchoring groups that could be from the same or different bipyridine 
ligands. On the other hand, the heteroleptic dyes can attach through the only two available carboxylic groups on 
its bipyridine ligand.  

\begin{figure}
 \centering
 	\includegraphics[width=\columnwidth]{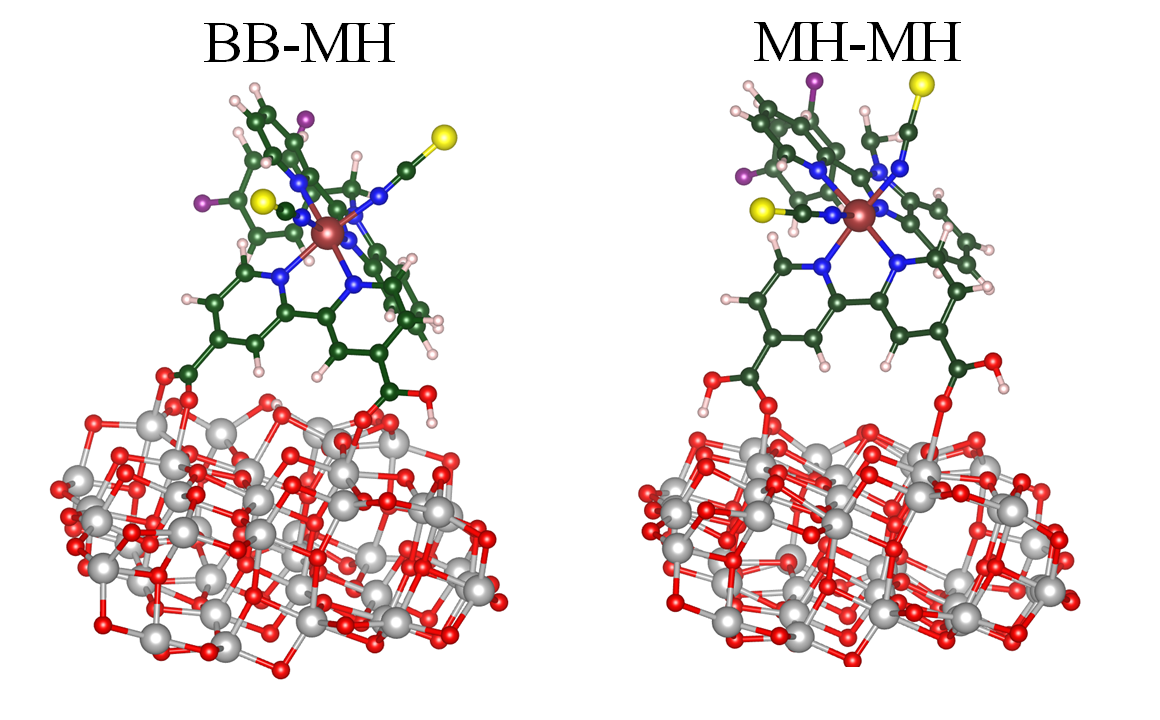}
 		\caption{Adsorption geometry of RD12 on the anatase (TiO$_2$)$_{38}$ cluster for two most stable BB-MH and MH-MH configurations.} 
 	\label{fig10}
 \end{figure}

In RD dyes, examining the distance between the nearest oxygen atoms on two carboxylic groups of the anchoring 
ligand, it turns out that the relative orientation, shown in figure~\ref{fig10}, 
has the best structural matching with the five-coordinated Ti surface atoms.
For this relative orientation, each of the carboxylic group can attach in one of the forms of bidentate-bridging 
(BB), protonated-monodentate (MH), or deprotonated-monodentate ester-type (M).\cite{martsinovich2012adsorption} The 
different combinations resulted 
from the two carboxylic groups constitutes the set of adsorption modes. Different adsorption modes would lead to 
different level alignments and absorption spectra for the combined dye/nanoparticle 
system.\cite{deangelis2008,sanchez2012a,sanchez2012b}  Our calculations show that the BB-MH 
combination is the first most stable adsorption mode (with highest binding energy) and MH-MH is the next one which 
are shown in 
figure~\ref{fig10}. Moreover, our calculations show that the optimization of the system in BB-M mode, in 
which the proton of carboxylic group is attached to the nearest surface oxygen, ends up to the BB-MH mode implying 
that the ester-type mode of adsorption is not stable. It should be mentioned that in reality the deprotonation 
degree of carboxylic groups highly depends on the pH and composition of the electrolyte 
solution.\cite{mosconi2012solvent}
Our calculations for the adsorption of RD5, RD12, and RD15 dyes show that adding fluorine atoms on ancillary ligand 
has no important effects on the dye-surface bond lengths ($\sim~0.01~\AA $) and binding energies ($\sim~0.1~eV 
$).    
 
\begin{figure}
 \centering
 	\includegraphics[width=\columnwidth]{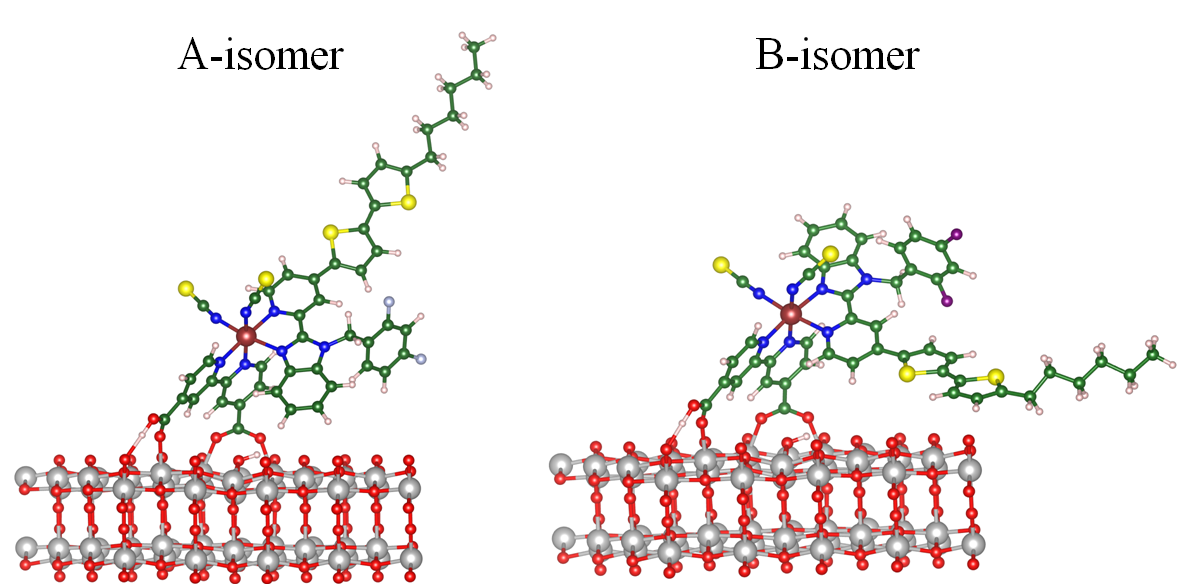}
 		\caption{Adsorption geometries of A- and B-isomers of RD18 on TiO$_2$ slab. As is seen, the B-isomer occupies a larger surface area which in turn leads to a significantly lower surface coverage. } 
 	\label{fig11}
 \end{figure}

As mentioned earlier, these RD dyes can be found in one of the two stereoisomeric forms, A- and B-isomer. For RD5, 
RD12, and RD15, the binding energy and surface coverage does not change significantly for the two stereoisomeric 
forms, whereas for RD18, because of its long hexylthiophene group, the surface coverage significantly decreases 
from A- to B-isomer (figure~\ref{fig11} ), and the B-isomer is more bound to the surface by 0.5~eV.  
This explains the experimental drop\cite{Huang2013} of dye-loading from 340~nmol/cm$^2$ for RD12 to 230~nmol/cm$^2$ for RD18.

For a better visualization of the relative positions of SCN ligands and TiO$_2$ surface in different combined RD/TiO$_2$ systems, we have made use of flat surfaces for TiO$_2$ nanoparticles in figure~\ref{fig14} of the next subsection. 

\subsection{Interaction of dye molecules with electrolyte components }
We know that: 

({\it i}) The open-circuit potential of a cell depends on the conduction band edge ($E_{\rm CB}$) and charge density ($n$) through\cite{ronca2013} 
\begin{equation}\label{eq11}
E_{F,n}=E_{\rm CB} + k_{\rm B}T \ln [n/N_c]
\end{equation}
where $E_{F,n}$ and $N_c$ are the quasi-Fermi level and the density of states of the semiconductor, respectively.  
$E_{\rm CB}$ strongly depends on the adsorption mode of the sensitizer, whereas $n$ depends on the recombination 
rate of injected electrons. All RD dyes have lower $V_{oc}$ compared to that of homoleptic N719 
dye,\cite{Huang2013} which can be explained to be as a result of the $E_{\rm CB}$ down-shift in heteroleptic dyes 
due to their adsorption modes.\cite{de2007} However, since the RD dyes have the same adsorption geometries, 
$V_{oc}$ is determined solely by the electron density of the semiconductor which, in turn, depends on the 
recombination 
rate of injected electrons;

\begin{figure}
 \centering
 	\includegraphics[width=0.6\columnwidth]{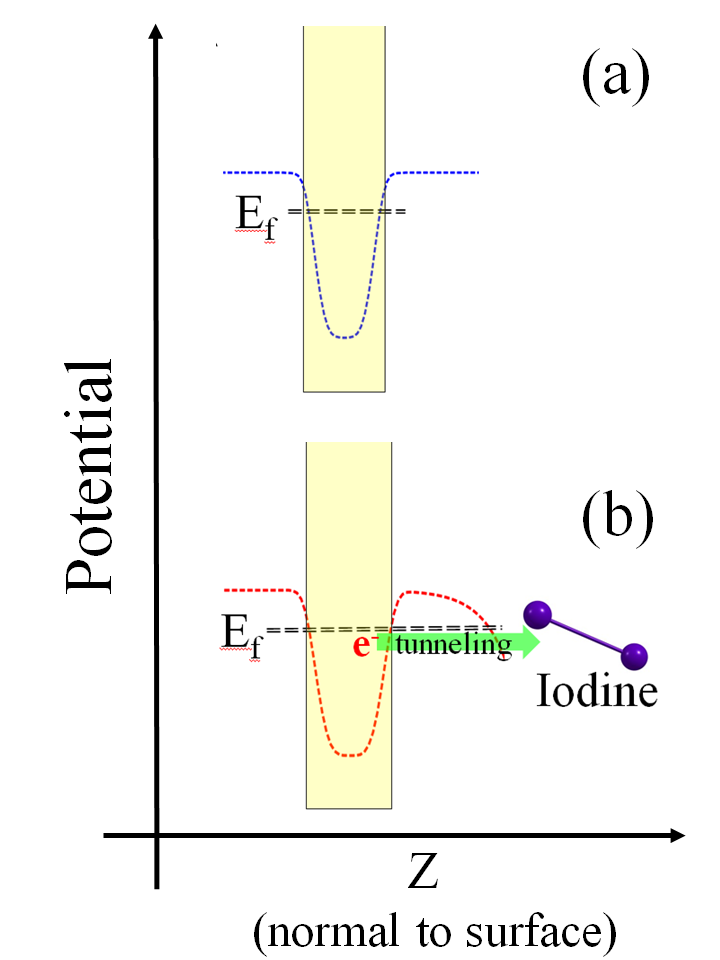}
 		\caption{Schematic representation of the electron tunnel from the surface. (a) and (b) represent the surface potential in the absence and in the presence of iodine molecule, respectively.}
 	\label{fig12}
 \end{figure}

({\it ii}) Among all electrolyte species, the iodine molecules (I$_2$) were shown to have main 
contribution in the recombination of electrons.\cite{Green2004} The rate of electron capture by these iodine 
molecules depends, firstly, on their concentration near the surface which, in turn, increases by  the concentration of the dye molecules,\cite{jeanbourquin2014} and secondly, depends on their relative 
orientation. Because of the $\sigma$-holes at the two ends of an I$_2$ molecule,\cite{Politzer13} an external 
electric field applies to the surface of the nanoparticle. This external field, in turn, modifies the confining 
electric potential near the surface in such a way that it becomes possible for an electron to escape from the surface 
via the tunnelling process (See figure~\ref{fig12}).
For two iodine molecules with the same centre of mass distance from 
the surface, the 
tunnelling rate becomes higher for the molecule with larger orientation angle (which is due to the smaller 
potential-barrier-width);   

({\it iii}) It has been shown that\cite{jean2014,azar2014} the attractive sites 
in a dye molecule (electron-rich sites) attract the I$_2$ molecules in the electrolyte to form a 
``dye$\cdots$I$_2$'' complex, which in turn, increases the recombination rate. 

Based on the discussions in ({\it i}), ({\it ii}), and ({\it 
iii}), to explain the observed variations in the recombination rates of different RD dyes, it is sufficient to investigate the possible orientations (relative to the surface) of the iodine molecules in ``dye$\cdots$I$_2$'' complexes.
However, since the sulphur atoms are the most attracting (electron-rich) sites\cite{kusama2011,jean2014} for halogen 
bonding in ruthenium complexes (see figure~\ref{fig5}), we consider only the halogen bonding with sulphur atoms 
in the dye molecules. 

 \begin{figure}
 \centering
 	\includegraphics[width=0.8\columnwidth]{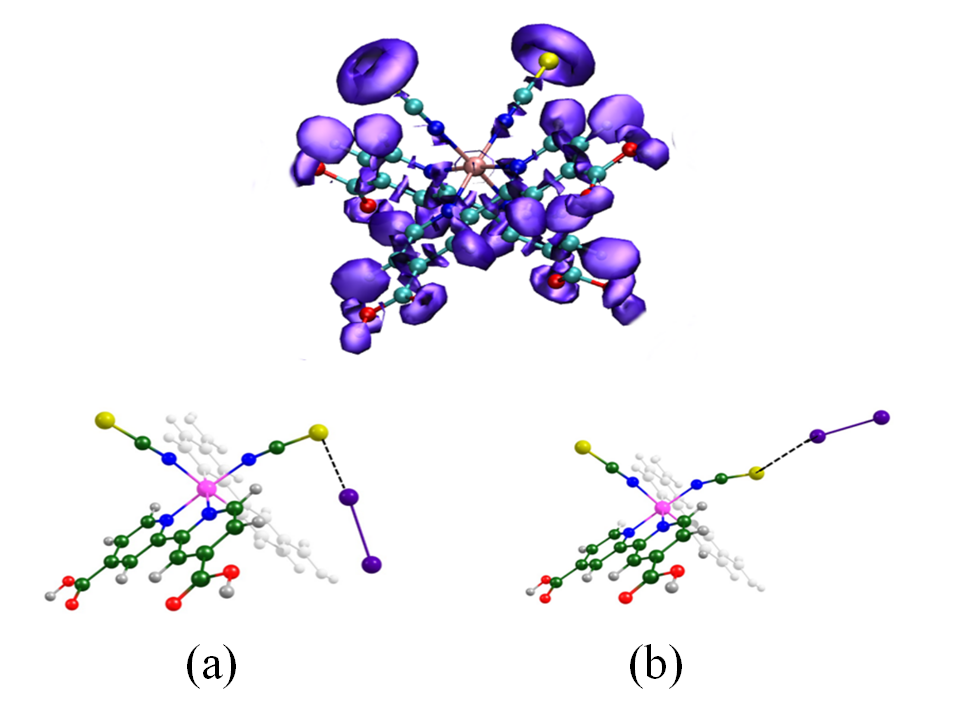}
 		\caption{Plot of electron localization function for isolated N3 (top), and different stable orientations 
 		of iodine molecule in N3$\cdots$I$_2$ complex (bottom). The complex for which the iodine bond is 
 		perpendicular to the SCN ligand (i.e., perpendicular to the torus axis in the top figure) (a), is more 
 		stable compared to the 
 		complex with parallel orientation (b). For a better representation, the ancillary ligand is shown as faded colourless.}  
 	\label{fig13}
 \end{figure}

To determine the geometry of halogen bonding to SCN ligand of N3 dye, we have considered two extreme relative orientations - ``perpendicular'' and ``parallel''- which was found to be stable configurations. The intermediate orientations reduce to one of the two mentioned stable configurations after optimization.
The configuration for which the iodine bond is perpendicular to SCN (''perpendicular'' orientation), was found to 
be more stable than the one with ``parallel'' orientation. This fact is readily understood by looking at the plot 
of electron localization 
function (ELF)\cite{Savin96,cauliez2010} shown in figure~\ref{fig13}. As is seen from figure~\ref{fig13}, the 
$\sigma$-hole of 
sulphur is along the SCN (i.e., along the torus axis) while those of iodine molecule lie at the two ends and 
therefore, the perpendicular orientation gives rise to a lower energy configuration.    

\begin{figure}
 \centering
 	\includegraphics[width=\columnwidth]{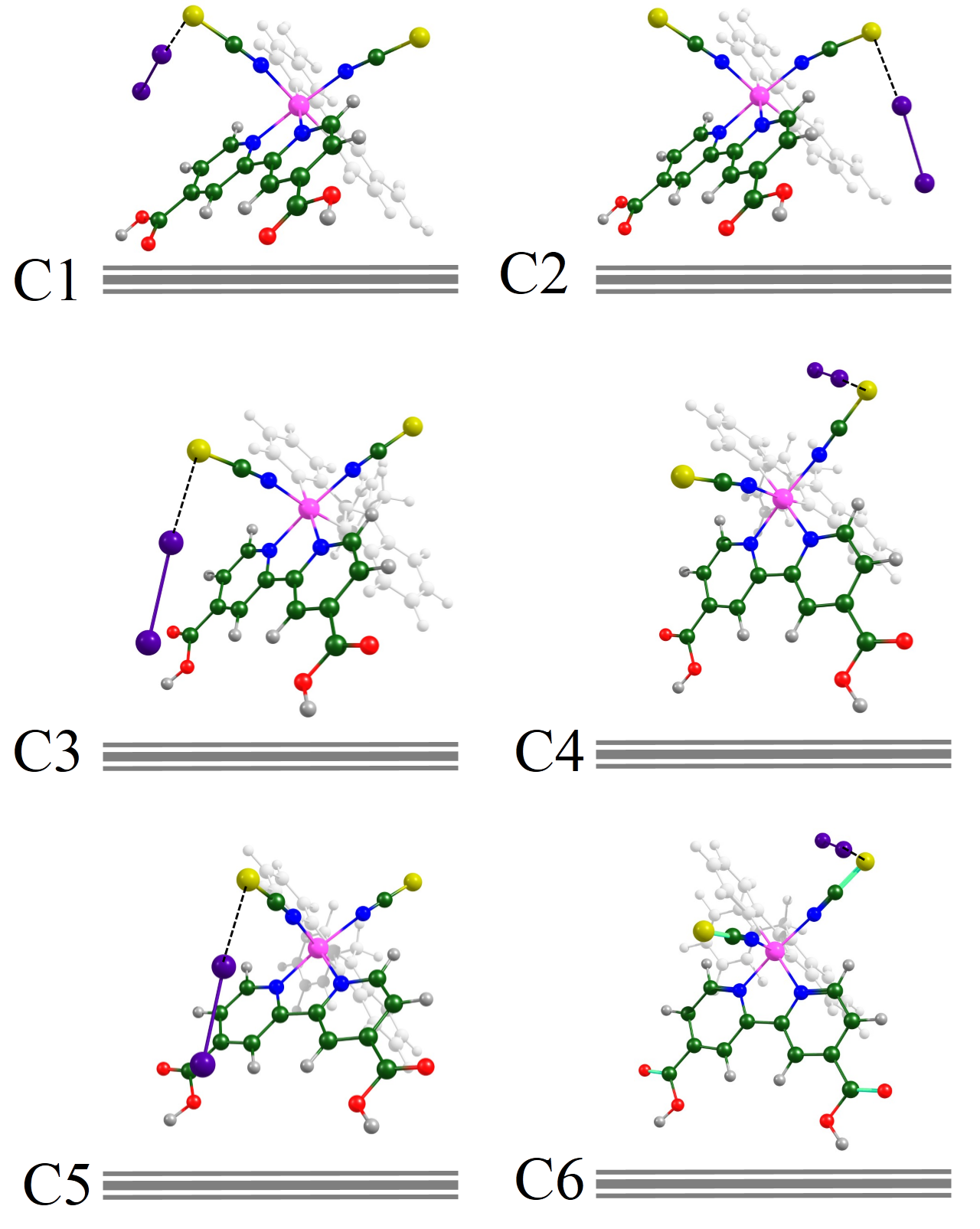}
 		\caption{Equilibrium configurations of ``dye$\cdots$I$_2$'' complexes for the more 
 		stable normal orientations. C1 and C2 refer to the bonding with the two different SCN ligands in N3. C3 
 		and C4 correspond to RD5, while C5 and C6 correspond to RD15. The orientations and distances relative to 
 		the surface of the iodine molecules in the two ligands are more or less the same for N3, while it is not 
 		the case for RD5 and RD15. As in figure~\ref{fig13}, the ancillary ligand is shown as faded colourless.
 		}  
 	\label{fig14}
 \end{figure}

Using the above fact that the most stable halogen bonding with SCN corresponds to the perpendicular orientations,  
 we have determined the equilibrium geometries of the complexes and shown the results in figure~\ref{fig14}.  
Our calculations show that within the perpendicular orientation to SCN, the iodine molecule can have different 
azimuthal directions (taking z-axis along SCN) with energy differences of at most $\sim 1$~kcal/mol. The azimuthal 
equilibrium position of the perpendicular iodine molecule is determined by the electrostatic interaction with the 
electrophile parts of ancillary and anchoring ligands.\cite{Politzer10}     

\begin{table}\label{tab4}
\caption{Bonding energies (in kcal/mol), bond lengths (in $\AA$), and transferred charges (in electron) for different configurations of dye$\cdots$I$_2$ complexes which are represented in figure~\ref{fig14}.}
\centering
\resizebox{\columnwidth}{!}
{
\begin{tabular}{lllllll} \hline
                  	& C1 & C2  & C3 & C4 & C5  &  C6   \\   \hline
	$\Delta E$   &  14.12   &  14.13     & 14.87     & 16.76    & 14.65     &  16.70     \\ 
	$d_{I...X}$     &  2.93     &  2.92      & 2.91      & 2.87      & 2.91       &   2.88      \\
	$\Delta q$	    &   0.31    &     0.31   &  0.32     & 0.36      & 0.32       &   0.36       \\  \hline
\end{tabular} 
} 
\label{tab4}
\end{table}

As shown in figure~\ref{fig14}, for N3, both SCN ligands have similar behaviours in bringing the iodine molecule 
near to the surface (C1 and C2). However, in RD dyes, one of the SCN ligands brings the iodine near to the surface 
(C3 and C5), while the other one keeps it far away from the surface (C4 and C6). This behaviour effectively halves 
(compared to N3 dye) the electron captures per dye molecule for RD dyes.

The bonding energies, bond lengths, and transferred charges of ``dye$\cdots$I$_2$'' complexes are tabulated in Table~\ref{tab4}. 
Concerning the RD dyes, the energy values in Table~\ref{tab4} show that the halogen bonding of iodine with 
that SCN ligand which is far from the surface, is stronger than the bonding with the closer one. Although the loading of RD dyes are about 
1.5 times larger than that of N3,\cite{Huang2012} because the number of attracting sites (near to surface) on RD 
dyes are halved, the overall effect is that the electron lifetime of RD dyes are greater than that of N3, in 
agreement with experiment.\cite{Huang2012}   
On the other hand, since the adsorption geometry and complex formation of RD dyes are more or less the same, the 
electron lifetime of these dyes is solely determined by their loading values. Therefore, RD15 with the lowest 
loading value has the largest electron lifetime (smallest recombination rate) consistent with 
experiment.\cite{Huang2012}

As to RD18, although the loading is about $60\%$ less than that of RDX (X=5, 12, 15) as discussed above, the number 
of attracting sites per unit area has been increased (because of the bithiophene group and its orientation relative to the surface in the stable B-isomer configuration) relative to 
that in RDX. This explains the relative increase in the observed recombination rate of thiophene contained 
dyes.\cite{Huang2013}  

\section{Conclusions}\label{sec4} 
In this work, we have employed DFT and TDDFT to investigate the electronic structure and absorption spectra of N3 and RD dyes both in vacuum and in DMF solvent. We performed calculations for N3 to use the results to describe the variations of the properties due to the structural modifications in RD dyes. 
The calculated results for orbitals' distributions show that for RD dyes, in contrast to N3, the distribution is not 
symmetric, and the HOMOs alternatively change the locations between two thiocyanate ligands whereas the LUMOs 
alternate between ancillary and anchoring ones. We have derived a formula based on DFT that can be used in conjunction with the MEP plots and orbital distributions over different atoms of a molecule to describe the level shifts in a solvent.
Examining the excitation corresponding to the first peak of UV/vis spectra showed that in our studied heteroleptic 
dyes, in contrast to 
N3, the charge is effectively transferred to the anchoring ligand, leading to higher $J_{sc}$ compared to the 
common homoleptic N3 dye. It should be mentioned that the PBE0 calculations lead to a better agreement of the absorption spectra with the experiment compared to other studied XC functionals. A simple formula in terms of the difference dipole moment vectors of the ground and 
excited states was written and used for illustration of charge transfer direction in an excitation process.
Finally, we have explained the different electron lifetimes observed in the RD dyes by investigating the adsorption geometries and the orientations of iodine molecules in different ``dye$\cdots$I$_2$'' complexes. 

\section*{Acknowledgement} 
Y.~T.~A. would like to thank Professor Eric Wei-Guang Diau and Dr. Wei-Kai Huang for the discussions on their experimental works. This work is part of research program in Theoretical and Computational Physics Group, AEOI.  

\footnotesize{
\bibliography{revised_payami} 
\bibliographystyle{rsc} 
}

\end{document}